\documentclass[pra,superscriptaddress,twocolumn]{revtex4}
\usepackage{amsfonts,amssymb,amsmath}
\usepackage[]{graphics,graphicx,epsfig}
\usepackage{amsthm}

\begin{document}

\title{Hidden variable problem for a family of continuously many spin 1 measurements}

\author{Pawe\l\ Kurzy\'nski}
\affiliation{Centre for Quantum Technologies,
National University of Singapore, 3 Science Drive 2, 117543 Singapore,
Singapore}
\affiliation{Faculty of Physics, Adam Mickiewicz University,
Umultowska 85, 61-614 Pozna\'{n}, Poland}

\author{Akihito Soeda}
\affiliation{Centre for Quantum Technologies,
National University of Singapore, 3 Science Drive 2, 117543 Singapore,
Singapore}

\author{Bart{\l}omiej Bzd\c{e}ga}
\affiliation{Department of Arithmetic Algebraic Geometry, 
Faculty of Mathematics and Computer Science, Adam Mickiewicz University,
Umultowska 87, 61-614 Pozna\'{n}, Poland}

\author{Dagomir Kaszlikowski}
\email{phykd@nus.edu.sg}
\affiliation{Centre for Quantum Technologies,
National University of Singapore, 3 Science Drive 2, 117543 Singapore,
Singapore}
\affiliation{Department of Physics,
National University of Singapore, 3 Science Drive 2, 117543 Singapore,
Singapore}

\begin{abstract}
We study a continuous set of spin 1 measurements and show that for a special family of measurements parametrized by a single variable $\theta$ the possibility of hidden-variable description is a discontinuous property.
\end{abstract}

\maketitle

\section{Introduction} 

Results of measurements on a three-level quantum system cannot be explained in a classical way. In quantum theory it is common to express fundamental exclusive events for such system as orthogonal unit vectors and the acclaimed Gleason's theorem \cite{G} states that in a three-dimensional space it is not possible to assign binary values 0 and 1 to vectors on a unit sphere, such that for all mutually orthogonal triples the value 1 is assigned to exactly one vector. In this case vectors that are assigned 1 correspond to the events that would be actually observed if the measurements were performed, hence deterministic description of quantum measurements would be possible. The subsequent Kochen-Specker theorem (KS) \cite{KS} states that it is possible to find finite subsets of unit vectors that do not allow for the 0-1 assignment discussed by Gleason. 

The result of KS inspired many scientists to look for minimal sets of vectors that do not admit the 0-1 assignment (see \cite{SO} and \cite{C1} for minimal proofs in dimensions 3 and 4) and set the foundations for the new field of research on quantum contextuality which studies the dependence of a measurement outcome on measurement context. In this paper we come back to the original idea by Gleason and investigate the possibility of an outcome assignment for a family of continuous subsets of unit vectors in three dimensions corresponding to a special class of spin 1 observables that is parametrized by a single parameter $\theta$. We find that within this family the ability of the outcome assignment strongly depends on the rationality of $\theta$ and is therefore a discontinuous property.

The general concept of contextuality that originated from the ideas of Gleason and KS can be explained more precisely in the following way. Imagine a measurement of a physical property $A$ that is compatible with two different sets of other physical properties. If the properties belonging to different sets are not co-measurable then $A$ can be measured either with the first or with the second set. In this case we say that these two sets provide a context for the measurement of $A$. In the simplest case, the two contexts can be provided by measurements $B$ and $C$, for which $[B,C]\neq 0$, but $[A,B]=[A,C]=0$. 

Other than purely academic interest, there are two important motivations behind our studies.
Although nowadays it is rather commonly accepted that hidden variable description of quantum mechanics is somehow unnatural, it is still important to show which quantum phenomena admit hidden variable description and which do not. This importance stems from the quantum computation approach, namely, it is crucial to know which quantum resources can be simulated with classical means. Another motivation is related to experimental imperfections due to instability of measuring devices. In a realistic measurement the setting of the apparatus varies between consecutive repetitions which effectively results in a continuous set of measurements centered around the original one \cite{Pitovsky,Meyer,Kent,CK,Breuer}. Here we show that for a special family of continuous sets of spin 1 observables parametrized by a single parameter $\theta$ an arbitrarily small variation of this parameter may cause a change in a description of a system from the one that admits an outcome assignment to the one that does not and vice versa.

\section{Method}

The mathematical method we are going to use in this work was introduced in \cite{KZ} to study continuous measurements in a quantum nonlocality scenario. Consider a set $Q=\{\langle A B \dots \rangle\}$ of expectation values of products of some compatible observables $A, B,\dots$. We assume that every element in this set can be derived using the standard quantum formula
\begin{equation}
\langle A B \dots \rangle = \text{Tr}\{\rho A B \dots\}, \nonumber
\end{equation}
where $\rho$ denotes the state of a quantum system. If it were possible to simulate $Q$ with a set of non-contextual hidden variables $\Lambda$, then for all elements in $Q$ these hidden variables would have to satisfy
\begin{equation}
\langle A B \dots \rangle=\int_{\Lambda} d\lambda \rho(\lambda) I(A,\lambda)I(B,\lambda)\dots, \nonumber
\end{equation}
where $I(A,\lambda)$ ascribes one of possible values of $A$ to an outcome of the measurement of $A$ given the hidden variable state is $\lambda\in\Lambda$ and $\rho(\lambda)$ denotes a probability distribution over $\Lambda$. 

Next, consider two vectors
\begin{equation}
|Q\rangle=\begin{pmatrix} \vdots \\ \langle A B \dots \rangle \\ \vdots \end{pmatrix}, \nonumber
\end{equation}
whose entries are given by all elements from $Q$, and
\begin{equation}
|\Lambda\rangle=\begin{pmatrix} \vdots \\ \int_{\Lambda} d\lambda \rho(\lambda) I(A,\lambda)I(B,\lambda)\dots \\ \vdots \end{pmatrix}, \nonumber
\end{equation}
whose entries correspond to a simulation of elements of $Q$ with $\Lambda$. If it were possible to simulate quantum mechanical expectation values with $\Lambda$ one would obtain
\begin{equation}
\langle \Lambda|Q\rangle=\langle Q|Q\rangle, \nonumber
\end{equation}
We are interested in non-classical scenarios, i.e., the ones that cannot be simulated by any non-contextual $\Lambda$, for which one finds
\begin{equation}
\langle \Lambda|Q\rangle < \langle Q|Q\rangle. \nonumber
\end{equation}
Moreover, we focus on continuous sets $Q$ for which the above non-classicality condition can be rewritten as 
\begin{eqnarray}
& &\int_{Q} dA dB \dots \left(\langle A B \dots \rangle \int_{\Lambda} d\lambda \rho(\lambda) I(A,\lambda)I(B,\lambda)\dots \right) \nonumber \\ 
&<& \int_{Q} dA dB \dots \langle A B \dots \rangle^2. \label{cond}
\end{eqnarray}

\section{Spin 1 observables} 

We consider operators $A_{\vec{k}}=2S^2_{\vec{k}}-I$, where $S_{\vec{k}}$ is a spin 1 operator for direction $\vec{k}$. These are degenerate binary operators with two eigenvalues of $+1$ and one eigenvalue of $-1$. Due to the fact that in the case of spin 1, the squares of spin operators for orthogonal directions $\vec{k}$ and $\vec{l}$ are compatible, so are the corresponding observables $A_{\vec{k}}$ and $A_{\vec{l}}$. In fact, it is possible to determine the values of any three operators $A_{\vec{k}}$, $A_{\vec{l}}$ and $A_{\vec{m}}$ for which the directions $\vec{k}$, $\vec{l}$ and $\vec{m}$ form a mutually orthogonal triple. Since all three operators mutually commute it is possible to represent all three in a diagonal form
\begin{equation}
A_{\vec{k}}=\begin{pmatrix} -1 & 0 & 0 \\ 0 & 1 & 0 \\ 0 & 0 & 1 \end{pmatrix}~A_{\vec{l}}=\begin{pmatrix} 1 & 0 & 0 \\ 0 & -1 & 0 \\ 0 & 0 & 1 \end{pmatrix}~A_{\vec{m}}=\begin{pmatrix} 1 & 0 & 0 \\ 0 & 1 & 0 \\ 0 & 0 & -1 \end{pmatrix}.
\nonumber
\end{equation}
Before we proceed to the main problem of this work, in the following two sections we consider simple examples of continuous sets of observables $A_{\vec{k}}$ to which we apply the condition (\ref{cond}).

\section{Single observables}

We start by considering a set of expectation values of single spin 1 observables $Q=\langle A_{\vec{k}} \rangle$, where $\vec{k}$ corresponds to all directions in three-dimensional space. The condition (\ref{cond}) yields
\begin{equation}
\int_{Q} d\vec{k} \langle A_{\vec{k}} \rangle \int_{\Lambda} d\lambda \rho(\lambda) I(\vec{k},\lambda) < \int_{Q} d\vec{k} \langle A_{\vec{k}} \rangle^2. \nonumber
\end{equation}
This case is trivial since the set $Q$ does not involve any context, i.e., each observable is measured alone. One of the immediate consequences is that it is not possible to verify that for all pairs of compatible observables the assigned outcome cannot be $-1$ for both of them. Therefore, it is possible to consider the hidden variable set $\Lambda$ which consists of all $\pm 1$ assignments to observables $A_{\vec{k}}$. For these hidden variables one can always find a distribution $\rho(\lambda)$ such that
\begin{equation}
\int_{\Lambda} d\lambda \rho(\lambda) I(\vec{k},\lambda) = \langle A_{\vec{k}} \rangle \nonumber
\end{equation}
and hence
\begin{equation}
\int_{Q} d\vec{k} \langle A_{\vec{k}} \rangle \int_{\Lambda} d\lambda \rho(\lambda) I(\vec{k},\lambda) = \int_{Q} d\vec{k} \langle A_{\vec{k}} \rangle^2. \nonumber
\end{equation}
It follows that the method considered here is not capable of refuting such a hidden variable model.

\section{Triples of observables}

Let us now consider all possible products of compatible triples $A_{\vec{k}}A_{\vec{l}}A_{\vec{m}}$. For all mutually orthogonal directions $\{\vec{k},\vec{l},\vec{m}\}$ one has
\begin{equation}
A_{\vec{k}}A_{\vec{l}}A_{\vec{m}}=-I, \nonumber
\end{equation}
Therefore, for all quantum states
\begin{equation}
\langle A_{\vec{k}}A_{\vec{l}}A_{\vec{m}} \rangle=-1. \nonumber
\end{equation}
The set $Q$ consists of expectation values of all such triples and the non-classicality condition (\ref{cond}) simplifies to
\begin{equation}\label{GHZ}
-\int_Q d\vec{k}d\vec{l}d\vec{m}\int_{\Lambda}d\lambda \rho(\lambda)I(\vec{k},\lambda)I(\vec{l},\lambda)I(\vec{m},\lambda) < 1. 
\end{equation}

This condition can be violated only if $I(\vec{k},\lambda)I(\vec{l},\lambda)I(\vec{m},\lambda)=-1$ for all directions $\{\vec{k},\vec{l},\vec{m}\}$ and all $\lambda$ in the support of $\rho(\lambda)$. Moreover, due to the compatibility constraint exactly one element in the product $I(\vec{k},\lambda)I(\vec{l},\lambda)I(\vec{m},\lambda)$ has to be equal to $-1$, while the other two have to be equal to $+1$. However, this is not possible as this problem is equivalent to the 0-1 assignment considered by Gleason. Therefore, one cannot construct any hidden variable set satisfying the compatibility criteria.

One can imagine slightly relaxed constraints on hidden variables, namely that one demands that $-1$ cannot be assigned to more than one element in the triple (but allows the possibility to assign $+1$ to all three observables). Nevertheless, this assignment will always lead to inequality (\ref{GHZ}), therefore even in this case the system manifests non-classical behavior.

\section{Family of pairs of observables}

Finally, we consider a case in which the set $Q$ consists of expectations values of pairs $A_{\vec{k}}A_{\vec{l}}$. However this time, directions $\vec{k},\vec{l}=(x,y,z)$ that define the set do not span the whole unit sphere. They are of the form
\begin{equation}
\left(\cos\varphi\sin\theta,\sin\varphi\sin\theta,\cos\theta\right), \nonumber
\end{equation}
where $\theta$ is the fixed angle by which each vector is tilted from the $Z$ axis. The compatibility of $A_{\vec{k}}$ and $A_{\vec{l}}$ is provided if the corresponding vectors $\vec{k}$ and $\vec{l}$ are orthogonal, which can only happen if $\theta\in [\pi/4,\pi/2]$. From now on we use the following notation for the elements of the set $Q$
\begin{equation}
Q_{\theta}=\{\langle A(\varphi)A(\varphi+\Delta)\rangle\}, \nonumber
\end{equation}
where $\varphi \in [0,2\pi)$. The constant $\Delta$ is chosen such that $A(\varphi)$ and $A(\varphi+\Delta)$ are compatible. It is easy to verify using simple vector algebra that 
\begin{equation}\label{delta}
\Delta=\arccos\left(-\cot^2\theta\right).
\end{equation}
Note that the five measurements of the pentagram inequality \cite{KCBS} are a discrete subset from the family of sets $Q_{\theta}$, for which $\Delta=\frac{4\pi}{5}$. 

The state of the system $\rho$ is taken to be the $-1$ eigenstate of the operator $A_{\vec{z}}$, hence all the expectation values are $\varphi$-independent and can be easily evaluated to be
\begin{equation}
\langle A(\varphi)A(\varphi+\Delta)\rangle=1-4\cos^2\theta = g(\theta). \nonumber
\end{equation}
In this case the condition (\ref{cond}) simplifies to
\begin{eqnarray}
-\int_{0}^{2\pi} d\varphi \int_{\Lambda} d\lambda \rho(\lambda) I(\varphi,\lambda)I\left(\varphi+\Delta,\lambda\right) < 2\pi |g(\theta)|, \label{iq} \\
\int_{0}^{2\pi} d\varphi \int_{\Lambda} d\lambda \rho(\lambda) I(\varphi,\lambda)I\left(\varphi+\Delta,\lambda\right) < 2\pi |g(\theta)|. \label{iq2}
\end{eqnarray}
The above splitting into two cases is due to the fact that $g(\theta)$ changes sign at $\theta=\pi/3$. The condition (\ref{iq}) is valid for $\theta\in[\pi/4,\pi/3)$, whereas (\ref{iq2}) is valid for $\theta\in[\pi/3,\pi/2]$.

\subsection{The case $\theta\in[\pi/4,\pi/3)$}

We first study the case $\theta\in[\pi/4,\pi/3)$. We focus on the left hand side of inequality (\ref{iq}). Note that due to the properties of operators $A(\varphi)$, the outcome ascribing functions $I(\varphi,\lambda)$ return $+1$ or $-1$, however $I(\varphi,\lambda)$ and $I(\varphi+\Delta,\lambda)$ cannot be both $-1$ at the same time. Moreover, we need to maximize the left hand side of (\ref{iq}) which implies that hidden variable states should give $I(\varphi+\Delta,\lambda)=-I(\varphi,\lambda)$ for sufficiently many values of $\varphi$. By assuming that all hidden variable states in $\Lambda$ are optimal in the above sense, we have
\begin{eqnarray}
&-&\int_{0}^{2\pi} d\varphi \int_{\Lambda} d\lambda \rho(\lambda) I(\varphi,\lambda)I\left(\varphi+\Delta,\lambda\right) \nonumber \\ \leq &-&\int_{0}^{2\pi} d\varphi  f(\varphi)f(\varphi+\Delta), \nonumber
\end{eqnarray}
where $f(\varphi)$ is an optimal $\pm 1$ function. Note, in case the above formula becomes an equality and 
\begin{equation}
-\int_{0}^{2\pi} d\varphi  f(\varphi)f(\varphi+\Delta) > 2\pi |g(\theta)|, \nonumber
\end{equation}
one can always supplement $\Lambda$ with non-optimal hidden variable states for which $I(\varphi,\lambda)I\left(\varphi+\Delta,\lambda\right)=1$ to warrant the reproducibility of quantum theoretical values.

We are going to prove the following three statements
\begin{itemize}
\item $\min \int_{0}^{2\pi} d\varphi  f(\varphi)f(\varphi+\Delta) = -2\pi$ if $\Delta$ is irrational multiple of $2\pi$;
\item $\min \int_{0}^{2\pi} d\varphi  f(\varphi)f(\varphi+\Delta) = -2\pi$ if $\Delta=2\pi \frac{p}{q}$, where $p$ and $q$ are co-prime integers and $q=2n$;
\item $\min \int_{0}^{2\pi} d\varphi  f(\varphi)f(\varphi+\Delta) = -\frac{2n-1}{2n+1}2\pi$ if $\Delta=2\pi \frac{p}{q}$, where $p$ and $q$ are co-prime integers and $q=2n+1$.
\end{itemize}

First, let us note that arguments $\varphi\in [0,2\pi)$ form a group $G$ with addition modulo $2\pi$. One may consider a subgroup generated by the element $\Delta\in G$, namely $\langle\Delta\rangle=\{0,\Delta,2\Delta,\dots\}$. Note, that if $\Delta$ is an irrational multiple of $2\pi$, the subgroup $\langle\Delta\rangle$ is infinite. Otherwise, when $\Delta=2\pi \frac{p}{q}$ (with $p$ and $q$ co-prime) the order of $\langle\Delta\rangle$ is $q$. Moreover, one can decompose $G$ into cosets with respect to $\langle\Delta\rangle$. Next, according to the axiom of choice there exists a set $A$ that contains exactly one representative element from each such coset. In this case every element of $G$ can be uniquely represented as $r+k\Delta$, where $r\in A$ and $k$ is an integer.

Let us start with the case of $\Delta$ being an irrational multiple of $2\pi$. We set $f(r+k\Delta)=(-1)^k$, which assures that the value of $f$ for any two arguments separated by $\Delta$ can never be $-1$ at the same time. Therefore, $\int_{0}^{2\pi} d\varphi  f(\varphi)f(\varphi+\Delta) = -2\pi$ which proves the first statement, since this is the smallest value one can get. Next, we consider $\Delta=2\pi \frac{p}{q}$ for $q=2n$. This time the set $A$ can be explicitely given as $A=[0,\frac{\pi}{n})$ and $k\in\{0,1,\dots,2n-1\}$. Due to the fact that $k$ can take an even number of different values we can once more set $f(r+k\Delta)=(-1)^k$. Following the previous arguments, we see that this implies the second statement. 

The case of $\Delta=2\pi \frac{p}{q}$ for $q=2n+1$ is more complicated since this time we cannot choose $f(r+k\Delta)=(-1)^k$. First, we note that $\sum_{k=-n}^n f(r+k\Delta)f(r+(k+1)\Delta)\geq -2n+1$ because the products $f(r+k\Delta)f(r+(k+1)\Delta)$ cannot be all equal to $-1$. At least one of them has to be equal to $+1$. We can again give explicitly $A=[0,\frac{2\pi}{2n+1})$ and $k\in\{-n,\dots,n\}$. We have
\begin{eqnarray}
& &\int_{0}^{2\pi} d\varphi  f(\varphi)f(\varphi+\Delta) =  \nonumber \\
& &\sum_{k=-n}^n \int_A dr  f(r+k\Delta)f(r+(k+1)\Delta) =  \nonumber \\
& &\int_A dr  \left( \sum_{k=-n}^n f(r+k\Delta)f(r+(k+1)\Delta) \right) \geq \nonumber \\
& &\int_A dr (-2n+1) = -\frac{2n-1}{2n+1}2\pi. \nonumber
\end{eqnarray}
What is left is to find an optimal function $f$. One can easily check that the function $f(r+k\Delta)=(-1)^{|k|+n+1}$  attains the lower bound of the above inequality, which proves the last statement.

It follows, that for $\theta\in[\pi/4,\pi/3)$ the measurement results that may not be explainable by hidden variable theory are only those corresponding to $\Delta=2\pi \frac{p}{q}$ with $q=2n+1$. In this case we can use the above results to rewrite the condition (\ref{iq})
\begin{equation}
\frac{2n-1}{2n+1} <  4\cos^2 \theta -1, ~~ \theta\in [\pi/4,\pi/3). \nonumber
\end{equation}
Moreover, using Eq. (\ref{delta}), simple algebra, trigonometry and writing explicitly $\Delta=2\pi\frac{p}{2n+1}$ one can obtain
\begin{equation}\label{p}
p >  \frac{2n+1}{2\pi}\arccos\left(-\frac{n}{n+1}\right). 
\end{equation}
On the other hand, $\theta\in [\pi/4,\pi/3)$ requires
\begin{equation}\label{p2}
\arccos(-1/3)/2\pi \approx 0.3 < \frac{p}{2n+1}\leq 0.5,
\end{equation}
which implies that $p\leq n$ and approximately $p> 0.6 n $. One can easily verify that except for $n=1$ the condition (\ref{p}) is always satisfied if one chooses $p=n$ and therefore hidden variable description of measurements can be refuted. For $n=1$, Eqs. (\ref{p}) and (\ref{p2}) can never be satisfied, hence this case admits hidden variables.

\subsection{The case $\theta\in[\pi/3,\pi/2)$}

Finally, let us consider $\theta\in[\pi/3,\pi/2]$. In this case we use the condition (\ref{iq2}), however it can be easily verified that the assignment corresponding to the uniform function $\tilde{f}(\varphi)=+1$ already violates this condition. Moreover, as in the previous case, the hidden variable simulation can be done by mixing $\tilde{f}(\varphi)$ with optimal $f(\varphi)$ from the last example. Therefore, the family $Q_{\theta}$ for $\theta\in[\pi/3,\pi/2]$ has hidden variable description.

\vspace{5mm}

\section{Discussion}

The result presented in the previous section shows that non-classical description of measurement families $Q_{\theta}$ is not a contiunuous property. In an arbitrarily small neighborhood of non-classical set $Q_{\theta}$ there always exists a set $Q_{\theta + \varepsilon}$ that admits hidden variable description. Apart from the fact that this behavior is somehow counterintuitive, since we are rather used to continuous behavior in physical phenomena, the above result may have implications for experimental scenarios \cite{Pitovsky,Meyer,Kent,CK,Breuer}. 

In our recent work \cite{us} we studied necessary conditions for the non-classicality of sets of quantum measurements. This non-classicality was defined by the lack of joint probability distribution for all measurements which is equivalent to the lack of hidden variable model \cite{Fine}. We found that a necessary condition for a set of measurements $S=\{A,B,\dots\}$ to reveal non-classicality is that there exists a subset $S'=\{A,C,E,\dots,X,Z\}\subset S$ whose elements obey cyclic commutation relation, i.e., 
\begin{equation}
[A,C]=[C,E]=\dots=[X,Z]=[Z,A]=0. \nonumber
\end{equation}
For three-level systems the cardinality of this subset has to be greater than four \cite{us}. Moreover, an experimental test based on $S$ has to allow one to evaluate the probabilities $p(A=a,C=c)$ for all pairs of commuting observables from $S'$. Finally, the non-classicality test may be based on the subset $S'$ alone. In the case considered in this work such subsets always exist if $\Delta$ is a rational multiple of $2\pi$. Moreover, it was shown that in case of projective measurements the cardinality of the non-classical set $S'$ has to be odd, which relates our present result to the previous research on discrete non-classical sets. Note, that the finite subgroup $\langle \Delta \rangle$ for $\Delta=2\pi\frac{p}{2n+1}$ generates cosets which correspond to non-classical sets studied in \cite{KCBS,CSW}. 

Finally, the non-classical set $Q$ for triples of observables is defined for all possible directions in a three-dimensional space. For any quantum state all expectation values in this set are the same and equal to $-1$. This cannot be reproduced by any hidden variable model. As we already mentioned, this set is equivalent to the one studied by Gleason. On the other hand, for the family $Q_{\theta}$ of pairs of observables the non-classical description strongly depends on the choice of quantum state. In this sense one may interpret our result for this family as state dependent Gleason's theorem (due to the continuous nature of the family). Note that for finite sets of measurement one can find both state-dependent \cite{KCBS} and state-independent \cite{SO, C1} non-classical sets.

We would like to thank Mateus Araujo and Marco Tulio Quintino for insightful discussions. This work is supported by the National Research Foundation and Ministry of Education in Singapore. P. K. is also supported by the Foundation for Polish Science.

\end{document}